\documentclass[12pt]{article}
\usepackage{latexsym} 
\usepackage{amssymb}  
\usepackage{amsmath} 
\usepackage{graphicx}
\usepackage{tabulary}
\usepackage{epsfig,wrapfig}
\usepackage[usenames]{color}
\usepackage[rflt]{floatflt} 

\begin{document}

\newcommand{\talk}[3]
{\noindent{#1}\\ \mbox{}\ \ \ {\it #2} \dotfill {\pageref{#3}}\\[1.8mm]}
\newcommand{\stalk}[3]
{{#1} & {\it #2} & {\pageref{#3}}\\}
\newcommand{\snotalk}[3]
{{#1} & {\it #2} & {{#3}n.r.}\\}
\newcommand{\notalk}[3]
{\noindent{#1}\\ \mbox{}\ \ \ {\it #2} \hfill {{#3}n.r.}\\[1.8mm]}
\newcounter{zyxabstract}     
\newcounter{zyxrefers}        

\newcommand{\newabstract}
{\clearpage\stepcounter{zyxabstract}\setcounter{equation}{0}
\setcounter{footnote}{0}\setcounter{figure}{0}\setcounter{table}{0}}

\newcommand{\talkin}[2]{\newabstract\label{#2}\input{#1}}                

\newcommand{\rlabel}[1]{\label{zyx\arabic{zyxabstract}#1}}
\newcommand{\rref}[1]{\ref{zyx\arabic{zyxabstract}#1}}

\renewenvironment{thebibliography}[1] 
{\section*{References}\setcounter{zyxrefers}{0}
\begin{list}{ [\arabic{zyxrefers}]}{\usecounter{zyxrefers}}}
{\end{list}}
\newenvironment{thebibliographynotitle}[1] 
{\setcounter{zyxrefers}{0}
\begin{list}{ [\arabic{zyxrefers}]}
{\usecounter{zyxrefers}\setlength{\itemsep}{-2mm}}}
{\end{list}}

\renewcommand{\bibitem}[1]{\item\rlabel{y#1}}
\renewcommand{\cite}[1]{[\rref{y#1}]}      
\newcommand{\citetwo}[2]{[\rref{y#1},\rref{y#2}]}
\newcommand{\citethree}[3]{[\rref{y#1},\rref{y#2},\rref{y#3}]}
\newcommand{\citefour}[4]{[\rref{y#1},\rref{y#2},\rref{y#3},\rref{y#4}]}
\newcommand{\citefive}[5]
{[\rref{y#1},\rref{y#2},\rref{y#3},\rref{y#4},\rref{y#5}]}
\newcommand{\citesix}[6]
{[\rref{y#1},\rref{y#2},\rref{y#3},\rref{y#4},\rref{y#5},\rref{y#6}]}

\begin{titlepage}



\vspace*{0.5cm}

\begin{center}
  {\Huge \bf Workshop on }\\[0.5cm]
  {\Huge \bf Precision Measurements of $\alpha_s$}\\[15mm]
  MPI Munich, Germany\\
  February 9 to 11, 2011\\[1cm] 
  
  {\em Editors}\\[1cm]
  {\bf Siegfried Bethke$^1$, Andre H.~Hoang$^{2}$, Stefan Kluth$^1$, \\
  Jochen Schieck$^3$, Iain W.~Stewart$^4$
  }\\[0.3cm]
  {\em $^1$Max-Planck-Institut f\"ur Physik (Werner-Heisenberg-Institut)%
, 80805 M\"unchen }\\
  {\em $^2$University of Vienna, Faculty of Physics,  A-1090
    Vienna}\\
  {\em $^3$Fakult\"at f\"ur Physik, Ludwig-Maximilians-Universit\"at M\"unchen,
    M\"unchen}\\
  {\em $^4$Center for Theoretical Physics, MIT, Cambridge, MA 02139}

\end{center}

\vspace*{1.0cm}

\begin{abstract}
  \baselineskip 1.5em These are the proceedings of the Workshop on Precision
  Measurements of $\alpha_s$ held at the Max-Planck-Institute for Physics,
  Munich, February~9-11, 2011.  The workshop explored in depth the determination
  of $\alpha_s(m_Z)$ in the $\overline {\rm MS}$ scheme from the key categories
  where high precision measurements are currently being made, including DIS and
  global PDF fits, $\tau$-decays, electro-weak precision observables and
  $Z$-decays, event-shapes, and lattice QCD.  These proceedings contain
  a short summary contribution from the speakers, as well as the lists of
  authors, conveners, participants, and talks.
\end{abstract}

\vspace*{3.5cm}

\begin{flushleft}
\small{ Supported by: MPI Munich, TUM/LMU Excellence Cluster,  University of
  Vienna
}
\end{flushleft}

\end{titlepage}

\begin{center}

{\em Authors and Speakers}\\[0.5cm] 

\small
{\bf S.~Aoki (U. Tsukuba), M.~Beneke (RWTH Aachen), S. Bethke (MPI), 
  J.~Bl\"umlein (DESY),  N.~Brambilla (TUM), S.~Brodsky (SLAC),
  S.~Descotes-Genon (LPT), J.~Erler (IF-UNAM), S.~Forte (U. Milano), 
  T.~Gehrmann (U. Z\"urich), C.~Glasman (U. Madrid),
  M.~Golterman (San Francisco State U.), 
  S.~Hashimoto (KEK), S.~Kluth (MPI), A.~Kronfeld (Fermilab), 
  J.~K\"uhn (KIT), P.~Lepage (Cornell), A.~Martin (Durham U.),
  V.~Mateu (MIT), S.~Menke (MPI), Y.~Nomura (UC Berkeley), 
  C.~Pahl (MPI), F.~Petriello (Argonne/Northwestern), A.~Pich (U. Valencia), 
  K.~Rabbertz (KIT), G.~Salam (CERN/Princeton/LPTHE), H.~Schulz (HU Berlin), 
  R.~Sommer (DESY), M.~Steinhauser (KIT), B.~Webber (U. Cambridge), 
  CP.~Yuan (Michigan State U.),  G.~Zanderighi (Oxford U.)
}\\[1cm]

{\em Conveners and Speakers without written contributions }\\[0.5cm]
\small
{\bf C.~Davies (U. Glasgow), Y.~Dokshitzer (U. Paris-6), 
  A.~H\"ocker (CERN), W.~Hollik (MPI), K.~Moenig (DESY),
  V.~Radescu (Heidelberg), B.~Reisert (MPI) }\\[1cm] 

{\em Additional Participants}\\[0.5cm]
\small 
{\bf R.~Abbate (MIT), H.~Abramowicz (MPI), J.~Arguin (LBL), C.~Bobeth (TUM-IAS),
  D.~Boito (U. Barcelona), D.~Britzger (DESY), V.~Chekelian (MPI), 
 G.~Duckeck (LMU), M.~Fickinger (U. Arizona), P.~Fritzsch (HU Berlin),
 M.~Gouzevitch (CERN), G.~Grindhammer (MPI), G.~Heinrich (MPI), 
 M.~Jamin (ICREA/U. Barcelona), R.~Kogler (DESY), A.~Lenz (TUM), A.~Levy (MPI), 
 G.~Luisoni (U.~Durham), P.~Mastrolia (MPI), W.~Ochs (MPI), S.~Peris
 (U. Barcelona), 
C.~Sturm (MPI), J.~Terron (U. Madrid), A.~Vairo (TUM), S.~Weinzierl (U. Mainz)
}\\[1cm]

\includegraphics[height=7.2cm]{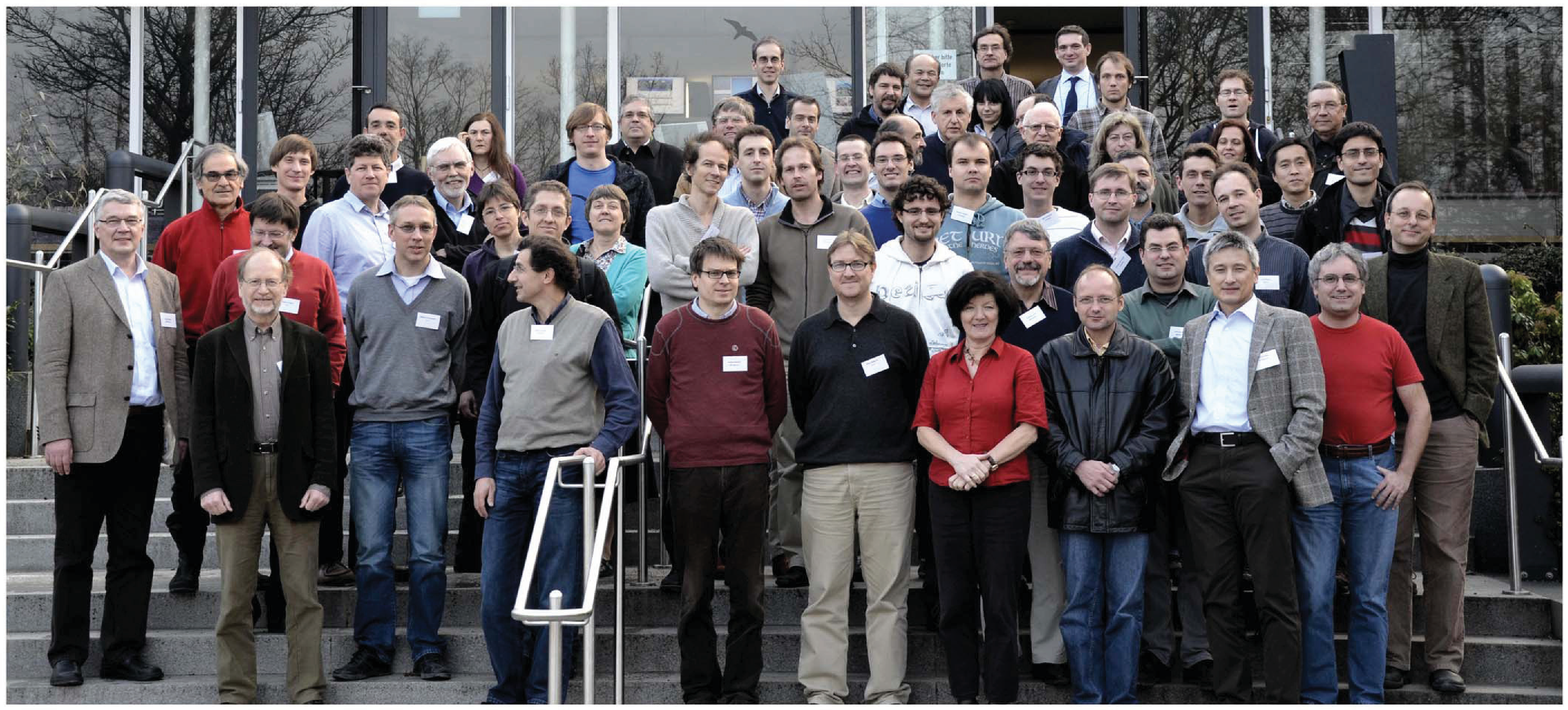} 
\end{center}

\setcounter{page}{2}

\newabstract

\section{Introduction}

The ``Workshop on Precision Measurements of $\alpha_s$'' was held at the Max
Planck Institute for Physics, February 9th through 11th of 2011. The meeting
brought together experts from several different fields where high precision
measurements of $\alpha_s(m_Z)$ are currently being made. Main goals of the
workshop were to facilitate discussion between the groups, and in particular to
give speakers the opportunity to explain details that one would normally not be
able to present at a conference, but which have an important impact on the
analyses. In each field the session was led off by a review speaker, followed
by more specialized talks, and was closed with a dedicated time period for
discussions.

There were 67 physicists who took part in the workshop, and 35 talks were
presented.  Slides as well as background reference materials are available on
the conference website
\begin{center}
    http://www.mpp.mpg.de/alphas
\end{center}
The sessions and talks in the workshop program were

\begin{itemize}
\itemsep -1.mm
\item Welcome
\vspace{-1ex}
\begin{itemize}
\itemsep -1.5mm
 \item  ``World Summary of $\alpha_s$ (2009) and beyond'', S.~Bethke
\end{itemize}
\item
$\alpha_s$ from Deep Inelastic Scattering and Global Fits
\vspace{-1ex}
\begin{itemize}
\itemsep -1.5mm
 \item ``Review of $\alpha_s$ Determinations from Jets at HERA'' by C.~Glasman 
 \item ``$\alpha_s$ from Deep-Inelastic Scattering: DESY Analysis'' by J.~Bl\"umlein
 \item ``CTEQ-TEA Parton Distribution Functions and $\alpha_s$'' by C.P.~Yuan
 \item ``$\alpha_s$ in MSTW Analyses'', A.~Martin
 \item ``Unbiased $\alpha_s$ from Global Fits: The NNPDF Approach'' by S.~Forte
 \item ``Hera PDF'' by B.~Reisert
 \item  Discussion Session on DIS and Global Fits, convened by V.~Radescu
\end{itemize}
\vspace{-1ex}
\item
Measurements of $\alpha_s$ from $\tau$ Decays
\vspace{-1ex}
\begin{itemize}
\itemsep -1.5mm
 \item ``$\alpha_s$ Determinations from Hadronic $\tau$ Decays'' by A.~Pich
 \item ``$\alpha_s$ from Contour Improved Perturbation Theory (CIPT)'' by
   S.~Descotes-Genon
 \item  ``Fixed Order Perturbation Theory (FOPT) Analysis'' by M.~Beneke
 \item  ``FOPT and CIPT in $\tau$ Decays'' by S.~Menke
 \item  ``Duality Violations in Hadronic $\tau$ Decays'' by M.~Goltermaan
 \item  ``Perturbative Input to $\tau$ Decays'' by J.~K\"uhn
 \item  ``Running and Decoupling of $\alpha_s$ at Low Scales'' by M.~Steinhauser
 \item Discussion Session on $\tau$ Decays, convened by A.~H\"ocker
\end{itemize}
\vspace{-1ex}
\item
$\alpha_s$ from $Z$ Decays and Electroweak Observables
\vspace{-1ex}
\begin{itemize}
\itemsep -1.5mm
 \item ``$\alpha_s$ in Electroweak Physics'' by J.~K\"uhn
 \item ``$\alpha_s$ with Global Analysis of Particle Properties (GAPP)'' by
   J.~Erler
 \item ``$\alpha_s$ from the Hadronic Width of the $Z$'' by K.~M\"onig
 \item Discussion Session on Elecroweak Analyses, convened by W.~Hollik
\end{itemize}
\vspace{-1ex}
\item
$\alpha_s$ from Event Shape Measurements
\vspace{-1ex}
\begin{itemize}
\itemsep -1.5mm
\item  ``Review of event-shape measurements of $\alpha_s$'' by G.~Salam
\item  ``$\alpha_s$ at NNLO and NNLA from (mainly) ALEPH data'' by T.~Gehrmann
\item  ``NNLO and Classic Power Corrections'' by B.~Webber
\item  ``$\alpha_s$ from Soft-Collinear Effective Theory analysis'' by V.~Mateu
\item  ``Five jets at LEP at NLO and $\alpha_s(m_Z)$'' by G.~Zanderighi
\item  ``Jet and Event Shape Observables at the LHC'' by K.~Rabbertz
\item  ``Monte Carlo Tuning with Professor'' by H.~Schulz
\item  ``Experimental Issues and
  Combination of Results'' by S.~Kluth              
\item  Discussion Session on Event Shapes, convened by Y.~Dokshitzer 
\item  MPI Colloquium, ``On the Other Side of Asymptotic Freedom'' by Y.~Dokshitzer     
\end{itemize}
\vspace{-1ex}
\item
Lattice Calculations and Quarkonia
\vspace{-1ex}
\begin{itemize}
\itemsep -1.5mm
\item ``Lattice QCD Calculations and $\alpha_s$'' by A.~Kronfeld 
\item ``The QCD Coupling from Lattice QCD (HPQCD)'' by G.P.~Lepage
\item ``$\alpha_s$ from Lattice QCD (PACS-CS)'' by S.~Aoki
\item ``$\alpha_s$ from Lattice QCD (JLQCD)'' by S.~Hashimoto
\item ``$\alpha_s$ from Lattice QCD (ALPHA)'' by R.~Sommer
\item ``$\alpha_s$ from Bottomonium'' by N.~Brambilla   
\item Discussion Session on Lattice QCD, convened by C.~Davies   
\end{itemize}
\vspace{-1ex}
\item Applications and Future of $\alpha_s$ Measurements
\vspace{-1ex}
\begin{itemize}
\itemsep -1.5mm
\item ``$\alpha_s$ for New Physics'' by Y.~Nomura 
\item ``The Strong Coupling and LHC Cross Sections'' by F.~Petriello
\item ``$\alpha_s$ at the International Linear Collider (ILC)'' by C.~Pahl 
\item ``The Principle of Maximum Conformality'' by S.~Brodsky
\item Final Discussion Session, convened by S.~Bethke
\end{itemize}
\vspace{-1ex}

\end{itemize}
\noindent This web proceedings represent a collection of extended abstracts and references
for the presentations, summarizing the most important results and issues. In
these writeups, unless otherwise indicated, the strong coupling $\alpha_s(\mu)$
is defined in the $\overline {\rm MS}$ scheme, and $\alpha_s(m_Z)$ is the
coupling with five light quark flavors.


\bigskip\bigskip

\noindent {\em Acknowledgments}\\[-5pt]

We would like to thank all participants for their effort to travel to Munich and
for making the Workshop on Precision Measurements of $\alpha_s$ a very stimulating
meeting. We cordially thank our secretary Mrs. Rosita Jurgeleit for her
excellent assistance, and the staff of the Max-Planck-Institute for their
invaluable support.

\bigskip

\noindent Munich, May 2011

\bigskip\bigskip\bigskip

\hspace*{10.cm} Siggi Bethke

\hspace*{10.cm} Andr\'e Hoang

\hspace*{10.cm} Stefan Kluth

\hspace*{10.cm} Jochen Schieck

\hspace*{10.cm} Iain Stewart

\bigskip\bigskip

%
%


\newpage

\section{Proceedings Contributions}

\vskip.1cm

\noindent\mbox{}\hfill{\bf Page}

\talk{{\bf S.~Bethke}}{Workshop Introduction}{abs:SiggiBethke_Intro}

\talk{{\bf C.~Glasman}}{Review of $\alpha_s$ Determinations from Jets at HERA}{abs:ClaudiaGlasman}

\talk{{\bf J.~Bl\"umlein}}{$\alpha_s(M_Z^2)$ in NNLO Analyses of Deep-Inelastic World Data}{abs:JohannesBluemlein}

\talk{{\bf CP.~Yuan}}{CTEQ-TEA Parton Distribution Functions and $\alpha_{s}$}{abs:CPYuan}

\talk{{\bf A.~Martin}}{$\alpha_S$ in MSTW Analyses}{abs:AlanMartin}

\talk{{\bf S.~Forte}}{$\alpha_s (M_Z)$  with  NNPDF Parton Distributions}{abs:StefanoForte}

\talk{{\bf A.~Pich}}{Tau Decay Determination of the QCD Coupling}{abs:AntonioPich}

\talk{{\bf S.~Descotes-Genon}}{$\alpha_s$ from $\tau$ Decays using Contour-Improved Perturbation Theory}{abs:SDescotesGenon}

\talk{{\bf M.~Beneke}}{Fixed-Order Analysis of the Hadronic $\tau$ Decay
  Width}{abs:MartinBeneke}

\talk{{\bf S.~Menke}}{FOPT and CIPT in $\tau$ Decays}{abs:SvenMenke}

\talk{{\bf M.~Golterman}}{Duality Violations in Hadronic $\tau$ Decays and the Value of $\alpha_s$}{abs:MaartenGolterman}

\talk{{\bf H.~K\"uhn}}{Perturbative Input to $\tau$ Decays}{abs:HansKuehn}

\newpage

\talk{{\bf M.~Steinhauser}}{Running and Decoupling of
  $\alpha_s$}{abs:MatthiasSteinhauser}

\talk{{\bf J.~Erler}}{$\alpha_s$ with GAPP}{abs:JensErler}

\talk{{\bf G.~Salam}}{Review of Event-Shape Measurements of $\mathbf{\alpha_s}$}{abs:GavinSalam}

\talk{{\bf T.~Gehrmann}}{$\alpha_s$ at NNLO(+NLLA) from (mainly) ALEPH Data}{abs:ThomasGehrmann}

\talk{{\bf B.~Webber}}{Next-to-Next-to-Leading Order and ``Classic'' Power Corrections}{abs:BrianWebber}

\talk{{\bf V.~Mateu}}{$\alpha_s$ from Thrust at N$^3$LL with Power Corrections}{abs:VicentMateu}

\talk{{\bf G.~Zanderighi}}{$\alpha_s$ from Five-Jet Observables at LEP}{abs:GiuliaZanderighi}

\talk{{\bf K.~Rabbertz}}{Jet and Event Shape Observables at LHC}{abs:KlausRabbertz}

\talk{{\bf H.~Schulz}}{MC tuning with Professor}{abs:HolgerSchulz}

\talk{{\bf S.~Kluth}}{$\alpha_S$ from Event Shapes in $e^+e^-$: Experimental Issues
and Combination of Results}{abs:StefanKluth}

\talk{{\bf A.~Kronfeld}}{Lattice QCD Calculations and \boldmath$\alpha_s$}{abs:AndreasKronfeld}

\talk{{\bf P.~Lepage}}{HPQCD: $\alpha_s$ from Lattice QCD}{abs:PeterLepage}

\talk{{\bf S.~Aoki}}{$\alpha_s$ from PACS-CS}{abs:SinyaAoki}

\newpage

\talk{{\bf S.~Hashimoto}}{$\alpha_s$ from JLQCD}{abs:ShojiHashimoto}

\talk{{\bf R.~Sommer}}{$\alpha_s$ from the ALPHA Collaboration}{abs:RainerSommer}

\talk{{\bf N.~Brambilla}}{$\alpha_{\rm s}$ Determination at the $Y(1S)$ Mass}{abs:NoraBrambilla}

\talk{{\bf Y.~Nomura}}{{\boldmath $\alpha_s$} for New Physics}{abs:YasunoriNomura}

\talk{{\bf F.~Petriello}}{The effect of $\alpha_S$ on Higgs Production at the LHC}{abs:FrankPetriello}

\talk{{\bf C.~Pahl}}{$\alpha_s$ from the ILC}{abs:ChristophPahl}

\talk{{\bf S.~Brodsky}}{The Principle of Maximum Conformality}{abs:StanBrodsky}

\talk{{\bf S.~Bethke}}{Workshop Conclusion}{abs:SiggiBethke_End}



\newabstract\label{abs:SiggiBethke_Intro}\input{SiggiBethke_Intro.in}

\newabstract\label{abs:ClaudiaGlasman}\input{ClaudiaGlasman.in}

\newabstract\label{abs:JohannesBluemlein}\input{JohannesBluemlein.in}

\newabstract\label{abs:CPYuan}\input{CPYuan.in}

\newabstract\label{abs:AlanMartin}\input{AlanMartin.in}

\newabstract\label{abs:StefanoForte}\input{StefanoForte.in}

\newabstract\label{abs:AntonioPich}\input{AntonioPich.in}

\newabstract\label{abs:SDescotesGenon}\input{SDescotesGenon.in}

\newabstract\label{abs:MartinBeneke}\input{MartinBeneke.in}

\newabstract\label{abs:SvenMenke}\input{SvenMenke.in}

\newabstract\label{abs:MaartenGolterman}\input{MaartenGolterman.in}

\newabstract\label{abs:HansKuehn}\input{HansKuehn.in}

\newabstract\label{abs:MatthiasSteinhauser}\input{MatthiasSteinhauser.in}

\newabstract\label{abs:JensErler}\input{JensErler.in}

\newabstract\label{abs:GavinSalam}\input{GavinSalam.in}

\newabstract\label{abs:ThomasGehrmann}\input{ThomasGehrmann.in}

\newabstract\label{abs:BrianWebber}\input{BrianWebber.in}

\newabstract\label{abs:VicentMateu}\input{VicentMateu.in}

\newabstract\label{abs:GiuliaZanderighi}\input{GiuliaZanderighi.in}

\newabstract\label{abs:KlausRabbertz}\input{KlausRabbertz.in}

\newabstract\label{abs:HolgerSchulz}\input{HolgerSchulz.in}

\newabstract\label{abs:StefanKluth}\input{StefanKluth.in}

\newabstract\label{abs:AndreasKronfeld}\input{AndreasKronfeld.in}

\newabstract\label{abs:PeterLepage}\input{PeterLepage.in}

\newabstract\label{abs:SinyaAoki}\input{SinyaAoki.in}

\newabstract\label{abs:ShojiHashimoto}\input{ShojiHashimoto.in}

\newabstract\label{abs:RainerSommer}\input{RainerSommer.in}

\newabstract\label{abs:NoraBrambilla}\input{NoraBrambilla.in}

\newabstract\label{abs:YasunoriNomura}\input{YasunoriNomura.in}

\newabstract\label{abs:FrankPetriello}\input{FrankPetriello.in}

\newabstract\label{abs:ChristophPahl}\input{ChristophPahl.in}

\newabstract\label{abs:StanBrodsky}\input{StanBrodsky.in}

\newabstract\label{abs:SiggiBethke_End}\input{SiggiBethke_End.in}

\end{document}